\documentclass[%
 reprint,
 amsmath,amssymb,
 aps,
prb,
]{revtex4-2}

\usepackage{graphicx}
\usepackage{dcolumn}
\usepackage{bm}
\usepackage[mathlines]{lineno}

\begin{document}


\title{Peculiarities Of High-Speed Dynamics Of Two- Photon Absorption In Si Nanowire Waveguides}


\author{Vadym Zayets}
\author{Siim Heinsalu}
\author{Akihiro Noriki}
 
\affiliation{National Institute of Advanced Industrial Science and Technology (AIST), Umezono 1-1-1, Tsukuba, Ibaraki, Japan}

\date{\today}

\begin{abstract}
We investigate the complete dynamical pathway of photon–electron interactions involved in two-photon absorption (TPA) in a silicon nanowire waveguide using three independent high-speed measurement techniques. These methods probe different stages of the process: nonlinear photon absorption, electron excitation from the valence to the conduction band, and free-carrier generation. According to the conventional model of TPA, these three processes should occur at identical rates. However, our measurements reveal significant discrepancies between them. The measured nonlinear photon absorption is more than twice the value required to account for the measured TPA transitions, indicating the presence of additional absorption pathways or nontrivial TPA dynamics. Furthermore, the number of measured TPA transitions substantially exceeds the measured free-carrier density, indicating that long-lifetime free carriers represent only a small fraction of the TPA-excited electrons, while the majority recombine rapidly back to the valence band on a timescale shorter than 13 ps. In addition, the three stages of the TPA pathway exhibit distinct saturation behaviors at different photon densities, further indicating that the TPA process in silicon is more complex than described by the conventional model. These findings provide new insight into the physical mechanisms governing TPA, suggesting the existence of multiple competing pathways for this optical transition. A major obstacle to a complete understanding of TPA is the unclear physical origin of the virtual midgap level. We identify three physically plausible origins for this level; however, further investigation is required to determine its true nature. The potential strategies for minimizing unwanted nonlinear losses in high-speed silicon photonic circuits, as well as for exploiting TPA in high-speed optical switching and photonic signal processing are investigated.
\end{abstract}

\keywords{two-photon absorption, optical transition, electron- phonon interaction, Si Photonics, all- optical data processing, optical computer, high-speed high-density data processing} 

\maketitle


\section{Introduction}

\subsection{TPA Limitations in Silicon Photonics}

Recent advances in hybrid photonic–electronic integration have enabled on-chip data transmission and processing at speeds approaching 10 Tbit/s by combining dense CMOS electronics with high-speed photonic components \cite{PolyWaveguide2026,Suda2025}. This integration unlocks exceptional performance in both speed and data density. However, as optical power and bit rates increase,the effect of the two-photon absorption (TPA) in silicon nanowire waveguides becomes increasingly prominent. As we demonstrate below, for short optical pulses in such waveguides, the nonlinear loss can substantially exceed the linear loss. Consequently,  TPA significantly degrades the bit error ratio (BER) and increases optical loss, thereby limiting both the achievable data rate and the maximum usable length of silicon waveguides.  These factors pose a major challenge for the further development of silicon photonics.  Therefore, it is essential to develop effective strategies to minimize unwanted nonlinear losses in high-speed silicon photonic circuits. Achieving this requires a detailed understanding of the underlying physical mechanisms governing TPA.

\begin{figure}[tb]
	\begin{center}
		\includegraphics[width=8.5 cm]{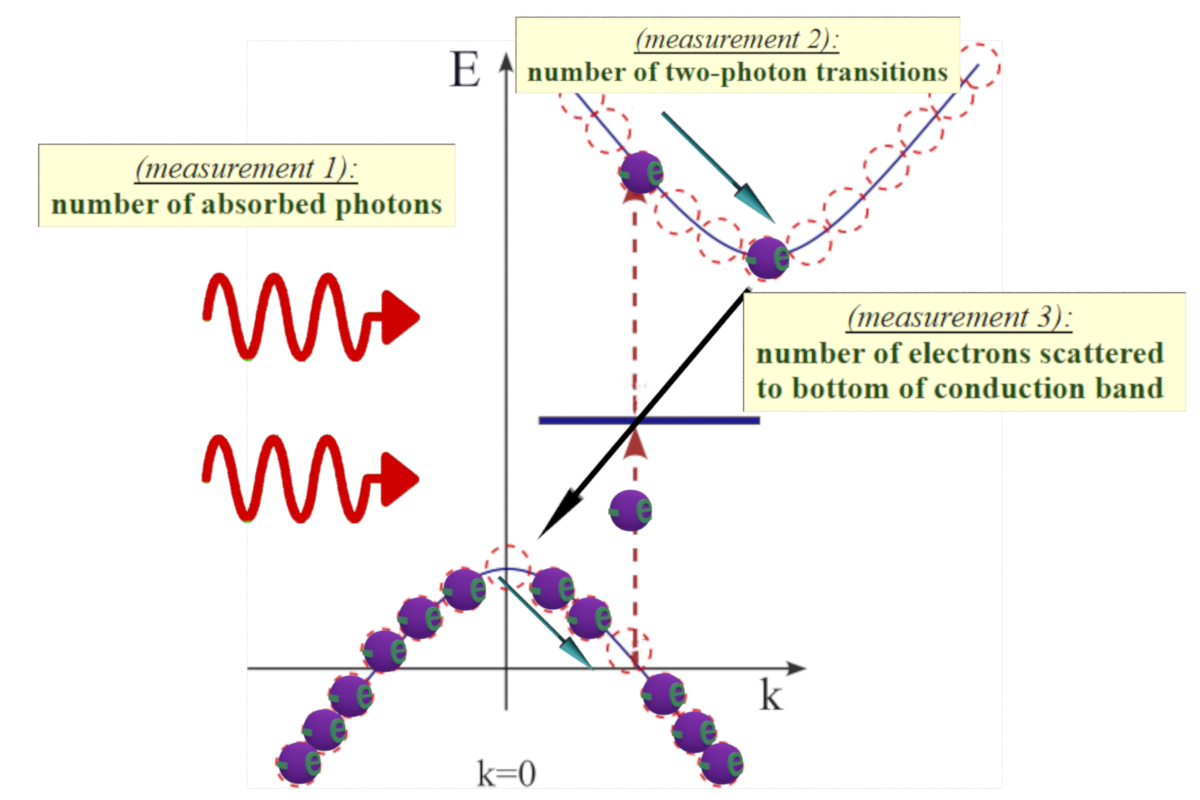}
	\end{center}
	\caption{
		{Tracing the band dynamics of the two-photon absorption process in Si. Three independent measurements evaluate the dynamics of three stages of the electron transition pathway. After absorbing the first photon, an electron is excited from the valence band to a virtual level (blue line). Upon absorption of the second photon, the electron is further excited to the conduction band. It then scatters down to the bottom of the conduction band and finally recombines with a hole at the top of the valence band via an indirect transition.
		} 
	}
	\label{fig:Dynamics} 
\end{figure}

\subsection{on TPA Origin: Mid-Gap Virtual States- Unavoidable Limitation or Functional Opportunity?}

TPA is a well-known and extensively studied phenomenon. The TPA mechanism is commonly described as involving a virtual energy level \cite{SiTPA_Kerr2007Bristow} within the bandgap (See Fig. \ref{fig:Dynamics}). In this model, an electron is first excited from the valence band to a virtual state by absorbing one photon and is subsequently excited to the conduction band by absorbing a second photon.

The existence of the virtual midgap level is generally regarded as an intrinsic material property that cannot be eliminated or modified. Consequently, it is widely believed that the only practical way to avoid TPA-related limitations in high-speed photonic systems is to use alternative materials that do not exhibit significant two-photon absorption, such as silicon nitride (SiN). SiN has a bandgap of approximately 4.0 eV, which exceeds twice the photon energy at a wavelength of 1.55 $\mu m$. As a result, the combined energy of two photons is insufficient to excite an electron from the valence band to the conduction band, making two-photon absorption energetically forbidden.  However, the combined energy of four photons exceeds the SiN bandgap, allowing four-photon absorption to occur. Although this process requires a substantially higher photon density (optical intensity) than TPA, it demonstrates that higher-order multiphoton absorption occurs alongside TPA \cite{multi2013, multiNSatur2001}. More importantly, the existence of higher-order multiphoton absorption indicates that virtual energy levels are not confined to the middle of the bandgap but are effectively distributed throughout the entire bandgap. 

 If it were possible to control the properties of the virtual midgap level-for example, through gate voltage modulation or controlled introduction of defects-the strength of TPA might also be controlled. Such control could provide two important benefits. First, unwanted speed limitations caused by TPA in silicon photonics could be reduced, enabling higher data processing speed and density. Second, modulation of TPA strength could be exploited as a mechanism for high-speed optical modulation and all-optical switching, transforming a traditionally limiting effect into a functional advantage.

Although controlling virtual levels in bulk silicon is challenging, it may be more feasible in ultra-thin silicon nanowire waveguides, where mid-gap states may naturally exist or be intentionally introduced at the silicon interface or by doping \cite{ZnOGa2026}. Mid-gap states at interface are common in semiconductors and are typically referred to as interfacial states. The bandgap in a semiconductor arises from the periodicity of the atomic lattice. When this periodicity is disrupted at an interface, interfacial mid-gap states can form. Because these interfacial states are intermixed with virtual mid-gap states, the strength of TPA may be modulated by altering the number and properties of surface states—for example, by applying an interfacial electric field or by modifying the crystal quality of the silicon interface.

All-optical switching based on TPA has been demonstrated in silicon nanowire waveguides with a reverse-biased p–i–n junction through modulation of free carriers \cite{pin_JuncRing2004,pin_JuncRing2005}. However, the required transport of free carriers makes such switching inherently slow, with switching times typically longer than 50 ps \cite{pin_JuncRing2005}. In contrast, direct TPA-based switching can occur on ultrafast timescales, potentially faster than 100 fs \cite{AutocorrelationXRay,Autocorrelation1992}. Control of the virtual level may enable the development of silicon-based optical modulators with switching times on the order of tens of femtoseconds.

\subsection{Unclear physical origin of virtual midgap level}

The physical mechanism responsible for the formation of mid-gap virtual states remains unclear. In the standard quantum- mechanical description \cite{Feynman}, virtual states are associated with processes in which short-lived transitions occur through a momentary violation of energy conservation law. An elementary particle may temporarily occupy a high-energy virtual level even when the total energy of the interacting particles is insufficient to reach that level. As long as the particle remains in this high-energy state for a duration shorter than the time allowed by the Heisenberg uncertainty principle, such a transition is permitted. The state is referred to as virtual because the particle cannot remain there permanently and can occupy it only for an extremely short time.

However, since TPA occurs in silicon but is absent in silicon nitride, the TPA process ultimately satisfies overall energy conservation. Therefore, the origin of the virtual level involved in TPA likely differs from the conventional interpretation of virtual states in quantum mechanics. 

\subsection{Inconsistency between Linear and Non-linear Losses}
\label{SectionLossLinear}

Another important question for understanding the TPA mechanism is why the virtual mid-gap  level leads to nonlinear optical loss while apparently not contributing to linear optical loss. For example, in the studied waveguides, the linear loss for the TE mode is approximately two times smaller than for the TM mode, whereas the nonlinear loss is at least two times larger for TE than for TM. In addition, we have measured both linear and nonlinear losses for samples with slightly different fabrication quality, yet no clear correlation between linear and non-linear losses has been observed.

If a mid-gap energy level exists-whether real or virtual-it should contribute proportionally to both linear and nonlinear absorption for the following reason. After excitation to this intermediate level, an electron has two possible pathways: it can either absorb a second photon and transition to the conduction band, contributing to TPA, or it can return to the valence band, contributing to linear absorption.  Consequently, the presence of a mid-gap level should fundamentally couple linear and nonlinear losses.

Both linear and nonlinear absorption would then be expected to scale with the density of states associated with this intermediate level. Consequently, some correlation between linear and nonlinear losses should be observed experimentally. However, no such correlation is found.

Furthermore, the probability of an electron returning to the valence band with photon emission is generally expected to be higher than the probability of absorbing a second photon and transitioning to the conduction band. Under this assumption, linear loss should exceed nonlinear loss, which again contradicts the experimental observations.

\subsection{Suggested Origins of Virtual Mid-gap Level}

Since the physical origin of TPA remains unclear, several interpretations of the nature of the virtual level are possible. One possible interpretation is that TPA “virtual” levels correspond to ordinary mid-gap quantum states, but with an extremely short electron lifetime. After excitation from the valence band, the electron may occupy a mid-gap level for a time shorter than the coherence time of holes and then return to its original state in the valence band emitting a photon. In this case, the emitted  photon remains fully coherent with the photons of the optical pulse, resulting in no linear optical loss. However, if during its short stay on the mid-gap level the electron interacts with a second photon and transitions to the conduction band, the number of photons in the optical pulse is reduced. This process leads to nonlinear optical loss.

The second interpretation assumes that the “virtual” mid-gap states do not exist prior to the arrival of the optical pulse but are instead created by the combined oscillating electric field of the photons within the pulse. Since the density of these states is proportional to the number of photons in the optical pulse, the optical loss they induce is inherently nonlinear.

Quantum states within the bandgap induced by the combined optical field of many photons are often referred to as field-dressed (or light-field-dressed) dynamic states \cite{LightDressedState1965,LightDressedState1973}. Similar to how a strong static electric field applied to an atom can shift and modify atomic energy levels (the Stark effect), a strong oscillating optical field can alter the electronic energy structure of a semiconductor.

The bandgap in a crystal originates from the periodic potential of the atomic lattice and the corresponding Bragg reflection of electron waves. A strong oscillating optical field perturbs this periodic potential. As the perfect periodicity becomes partially disrupted, the optical field can modify the band structure and enable the formation of quantum states within the bandgap.

The third interpretation assumes a mechanism of optical transition that is fundamentally different from a conventional two-particle optical process. In a standard interaction, a single photon interacts with a single electron. In contrast, two-photon absorption can be viewed as a three-particle interaction, in which two photons simultaneously interact with one electron. In this picture, the electron is excited directly from the valence band to the conduction band without requiring the presence of a “virtual” mid-gap state.

In this case, the existence of a virtual mid-gap level is not required to explain TPA. Since the transition is determined solely by the initial and final states in the valence and conduction bands, mid-gap states should not significantly influence the TPA process. Consequently, external modulation or suppression of TPA through control of mid-gap states would not be expected to be possible.

The fourth interpretation is based on interband electron tunneling from the valence band to the conduction band driven by the combined oscillating electric field of the optical pulse. This mechanism also does not require the presence of a virtual intermediate level. This mechanism is discussed in detail in Section \ref{sectionTPAvsKFR}. Since interband tunneling is primarily associated with ultrashort optical pulses, its contribution to the present measurements is expected to be negligible. Nevertheless, because the full range of interband tunneling phenomena is not yet completely understood, the possibility of a contribution from this mechanism cannot be entirely excluded and should be kept in mind when interpreting the results.

\subsection{Towards Understanding the TPA Mechanism: Tracing the Complete Pathway of TPA Dynamics}

To clarify the true mechanism of TPA, it is essential to trace the full dynamics of the electron-photon interaction, from excitation of an electron from the valence band to the conduction band to its subsequent relaxation back to the valence band. In particular, one must compare the number of photons absorbed nonlinearly with the number of electrons excited from the valence band to the conduction band by TPA, and subsequently with the number of free carriers that result from the relaxation of TPA-excited electrons and holes to the bottom of the conduction band and the top of the valence band.

For example, if the number of TPA-excited electrons does not match the number of nonlinearly absorbed photons, this would indicate either the presence of additional nonlinear processes contributing to the observed dynamics (e.g. interband tunneling), or that the TPA pathway involves an intermediate (virtual) state from which the process can be interrupted - allowing the electron to return to the valence band without reaching the conduction band.

Furthermore, if the number of free carriers is smaller than the number of TPA- excited electrons and holes, this would suggest that some electrons, after being excited to the conduction band, recombine rapidly back to the valence band. Tracing this recombination pathway could provide important insight into the underlying TPA mechanism.

An additional advantage of this approach is that the direct comparison of transition rates at the three stages may help distinguish between contributions from TPA and interband tunneling. Interband tunneling contributes only to Steps 1 and 3, whereas it does not contribute to Step 2. In contrast, TPA contributes to all three stages of the process. Therefore, differences in the measured transition rates may provide insight into the relative contributions of these two mechanisms (see Section \ref{sectionTPAvsKFR} for details).

The complete dynamical pathway of the photon- electron interaction in TPA within silicon has never been systematically investigated. The most frequently studied stage is free- carrier absorption, typically examined by probing pulse- excited free carriers using a continuous-wave (CW) beam \cite{SiReviewDekker2007,freeCarrirRelaxBrazil2018,freeCarrirRelaxChina2007} or using a pump- probe measurement \cite{PumpProbeWaveguide2010}. The two- photon absorption step itself has also been studied \cite{AISTSuda2011,Polarization2021,pulseRieger2004,pulse_NEC_2005,pulseZDinu2003,CWprobe2005} , usually by probing pulse-induced TPA with a delayed optical pulse.

Moreover, the conversion efficiencies corresponding to the different dynamical stages of TPA have never been compared comprehensively—not for all three steps, and in most cases not even for two. Data reported across different studies are difficult to compare due to variations in waveguide dimensions, fiber-to-waveguide coupling efficiencies, and measurement methodologies.

In this work, we measure and directly compare, for the first time, the conversion efficiencies across the full cycle of TPA dynamics. This approach enables us to address several important questions regarding the intrinsic mechanism of TPA. First, we examine whether all nonlinearly absorbed photons lead to electron transitions from the valence band to the conduction band, or whether only a fraction of these photons contribute to TPA. Second, we investigate whether all electrons excited to the conduction band subsequently relax to the conduction-band minimum, or whether additional relaxation and recombination pathways are involved.

\section{Experiment. Evaluation of photon dynamics}

Two-photon absorption was investigated in standard silicon nanowire waveguides with cross-sectional dimensions of 430 nm × 220 nm and a length of 5 mm, which are standard for  use in commercial silicon photonic circuits. The TE mode was studied, with measured linear propagation loss of 1.15 dB/cm.

Light coupling was achieved using lensed polarization-maintaining fibers. Spot-size converters were used at both facets minimizing the coupling loss to 2.2 dB per facet. Converters with tip widths of 130 nm, 140 nm, and 150 nm were tested. No dependence of the measured nonlinear loss on the converter width was observed, indicating a negligible contribution of the spot-size converters to the nonlinear loss measurements.

\begin{figure}[tb]
	\begin{center}
		\includegraphics[width=8.6 cm]{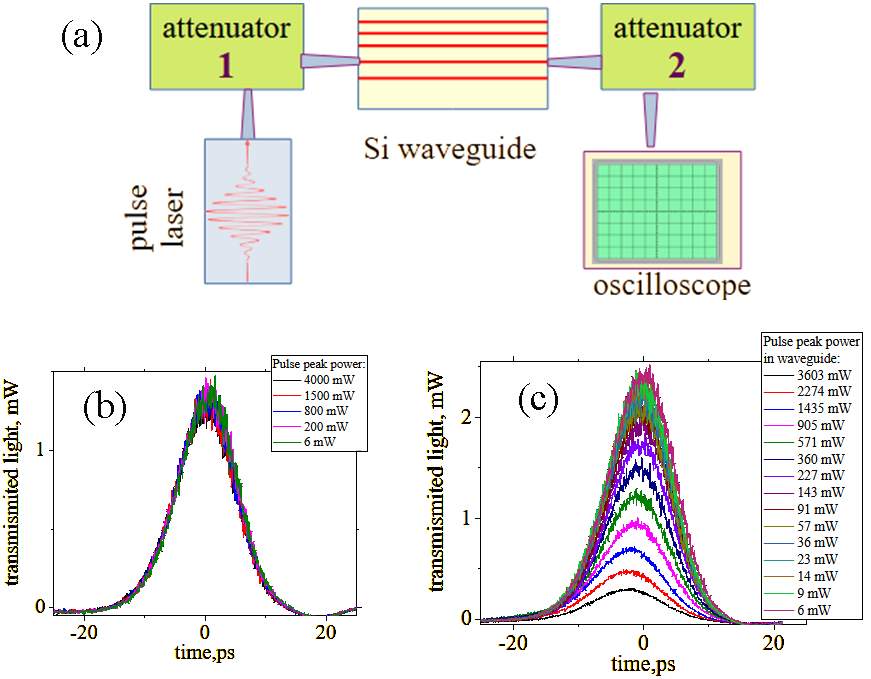}
	\end{center}
	\caption{
		{(a) Experimental setup for measuring the number of nonlinearly absorbed photons. Detected pulse dynamics (b) when the Si waveguide is bypassed, and (c) when the Si waveguide is inserted. The parameter in (b) is the pulse peak power after the first attenuator, while in (c) it is the pulse peak power inside the Si waveguide.
		} 
	}
	\label{fig:Exp1} 
\end{figure}

Figure \ref{fig:Exp1}(a) shows the proposed setup for measuring the number of nonlinearly absorbed photons. The setup is designed to measure the nonlinear loss of a single optical pulse. Although this type of measurement is widely used in studies of nonlinear optical phenomena, the experimental configuration was substantially modified to achieve the required accuracy and to obtain new information about the underlying absorption processes.

The main challenge of this measurement is the need to detect very small changes in pulse intensity with high precision while varying the pulse power over a wide dynamic range. If any component in the optical path, other than the silicon waveguide itself, exhibits even a weak nonlinear response, significant systematic errors may be introduced, potentially obscuring the true measurement results. The high-speed photodetector used in the oscilloscope is particularly known for having a very narrow linear dynamic range.

To overcome this limitation, we designed a setup with two variable attenuators, with the Si waveguide placed between them. The attenuators are computer-controlled so that when the attenuation of the first attenuator is changed, the attenuation of the second is automatically adjusted to keep the total attenuation constant. This configuration allows the pulse peak power inside the Si waveguide to be varied while keeping the total linear loss in the optical path unchanged.

Figure \ref{fig:Exp1}(b) shows the detected pulse dynamics vs the pulse intensity when the Si waveguide is bypassed. As expected, the detected peak power remains unchanged, confirming that the measurement system operates within the linear regime.

Figure \ref{fig:Exp1}(c) shows the detected pulse dynamics when the Si waveguide is inserted. In this case, the detected pulse intensity decreases significantly as the pulse power inside the waveguide increases, indicating substantial nonlinear loss in the Si nanowire waveguide. 

To evaluate the nonlinear photon loss (NPL), each curve in Fig. 1(c) was fitted with a Gaussian function, and the NPL (in dB/cm) was calculated as:

\begin{equation}
	Loss_{NPL}=\dfrac{10 \cdot log_{10}(A/A_0)}{L}
	\label{loss_photon}
\end{equation} 

where $A$ is the amplitude obtained from the Gaussian fit, $A_0$ is the amplitude at the lowest pulse intensity (for which the nonlinear loss is negligible), and $L$ is the waveguide length (5 mm).

The number of photons in an optical pulse is directly proportional to the pulse energy. Since only negligible pulse broadening was observed  (less than 1 ps at maximum- used pulse intensity), the pulse energy remains proportional to the peak pulse intensity A. Consequently, the rate of nonlinear loss the pulse photons $Rate_{NPL}$ can be evaluated as

\begin{equation}
	Rate_{NPL}=\dfrac{A_0-A}{A_0} \frac{1}{L}=\dfrac{1-A/A_0}{L}
	\label{rate_photon}
\end{equation}

\begin{figure}[tb]
	\begin{center}
		\includegraphics[width=8.6 cm]{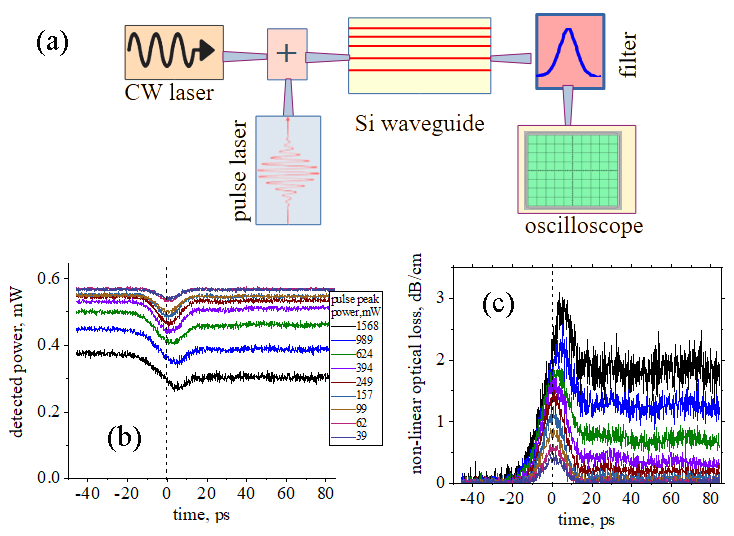}
	\end{center}
	\caption{
		{(a) Experimental setup for measuring the rate of two-photon absorption and the optical loss due to free-carrier absorption.(b) Output dynamics detected by the oscilloscope for different pulse peak powers inside the Si waveguide.(c) Temporal dynamics of the nonlinear loss evaluated from the data in (b). The dashed line indicates the center position of the pump pulse.
		} 
	}
	\label{fig:Exp2} 
\end{figure}

Figure \ref{fig:Exp2}(a)  shows the experimental setup used to measure both the rate of two-photon absorption and the optical loss due to free-carrier absorption. Similar setups were used to measure TPA \cite{CWprobe2005}  and FCA \cite{freeCarrirRelaxChina2007,freeCarrirRelaxBrazil2018}.  A pulse laser ($\lambda$=1553 nm, pulse width 13 ps ) was combined with a continuous-wave (CW) laser and coupled into the Si waveguide. After propagation through the waveguide, the pulse wavelength was filtered out using a 5-nm optical filter tuned to the CW laser wavelength. As a result, only the CW light reached the oscilloscope.

The polarization of both the pulsed pump and the CW probe was TE. Two CW wavelengths were tested: one close to the pulse wavelength ($\lambda=1559 nm$) and another further away  ($\lambda=1530 nm$).The results are nearly similar for these cases. Unless otherwise stated, all graphs shown in the paper correspond to the waveguide with a 140-nm-wide converter, measured at $\lambda=1559 nm$.

The pulse modulates the CW light through nonlinear absorption as shown in Fig. \ref{fig:Exp2}(b). The maximum absorption occurs at the temporal position of the pulse peak. Figure \ref{fig:Exp2}(c) shows the temporal dynamics of the nonlinear loss extracted from the data in Fig. \ref{fig:Exp2}(b), calculated as

\begin{equation}
	Loss_{CW}(t)=\dfrac{10 \cdot log_{10}(I(t)/I_{CW})}{L}-Loss_{before}
	\label{loss_dynamics}
\end{equation} 

where $I(t) $ is the detected instantaneous power (Fig. \ref{fig:Exp2}(b)),  $ I_{CW} $is the detected power in the absence of the pulse ( $\approxeq5.8 mW$ for data of Fig. \ref{fig:Exp2}(b)) and $Loss_{before}$ is the loss before arrival of the pulse (averaged from -40 to -20 ps).

The nonlinear loss shown in Fig. \ref{fig:Exp2}(c) consists of two distinct components. The first component closely follows the temporal shape of the pump pulse (see Fig. \ref{fig:Exp1}(c)), with a peak centered at the pulse maximum. This contribution arises from two-photon absorption, corresponding to the quantum transition of electrons from the valence band to the conduction band.

After the pump pulse has passed, a nearly constant residual loss remains. This component is attributed to free-carrier absorption (FCA), which originates from the Ohmic loss associated with the motion of electrons accumulated at the bottom of the conduction band and holes accumulated at the top of the valence band under the oscillating optical field.

Because silicon is an indirect-bandgap material, the free-carrier lifetime is relatively long ($\approxeq50 ns$). Therefore, within the 40 ps time window of our measurement, any temporal variation of the FCA appears negligible. The relaxation dynamics of free carriers in Si waveguides are complex\cite{freeCarrirRelaxChina2007,freeCarrirRelaxBrazil2017}  and have been experimentally studied  in Refs. \cite{freeCarrirRelaxChina2007,freeCarrirRelaxBrazil2018} using a similar experimental setup to that shown in Fig. \ref{fig:Exp2}(a), but with longer pump pulses and over a much longer timescale in the millisecond range.

As seen in Fig. \ref{fig:Exp2}(b), a nonlinear loss is present even before the arrival of the pump pulse. This residual loss is caused by free carriers accumulated from preceding pulses. The repetition period of the pulse laser is 52 ns, which is not long enough to allow complete relaxation of the free carriers. As a result, carriers accumulate over successive pulses. To eliminate the influence of this accumulated background, we accounted for it in Eq. (\ref{loss_dynamics}) by subtracting the pre-pulse baseline contribution, $Loss_{before}$.

\begin{figure}[tb]
	\begin{center}
		\includegraphics[width=6.5 cm]{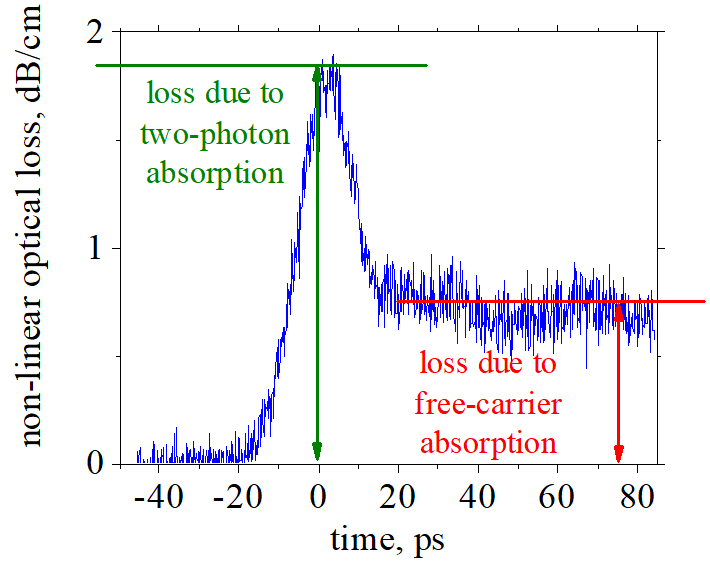}
	\end{center}
	\caption{
		{Evaluation of TPA and FCA losses from non-linear loss of CW light
		} 
	}
	\label{fig:FigLossTPApFCA} 
\end{figure}

The figure \ref{fig:FigLossTPApFCA} illustrates how the TPA and FCA losses were evaluated. It is important to note that the FCA was determined immediately after the two-photon transition was completed, before any significant free-carrier relaxation could occur, ensuring that measured free-carrier is all free- carrier generated by two-photon transition.

The rate of photon loss in the pulse due to TPA was evaluated from the measured TPA-induced loss of the CW probe. This type of CW loss occurs when one CW photon interacts with one photon from the pulse, resulting in an electron transition from the valence band to the conduction band. For this reason, the temporal profile of the TPA-induced CW loss follows a Gaussian shape that replicates the pulse envelope. Consequently, the TPA-induced loss $Loss_{TPA}$  is determined by the peak of the Gaussian distribution, indicated by the green curve in Fig. \ref{fig:FigLossTPApFCA}.

The rate of absorbed pulse photons $Rate_{TPA}$ due to TPA transitions is calculated as follows. The number of CW photons at input and output of Si waveguide is calculated from  $Loss_{TPA}$ as
\begin{equation}
		\frac{N_{out}}{N_{in}}=10^{-Loss_{TPA} \cdot L /10}\\
	\label{PhotonCW_TPA_CW}
\end{equation}

where $N_{in}$ and $N_{in}$ are numbers of CW photons at input and output of Si waveguide. The rate of CW photon loss $Rate_{TPA,cw}$ due to TPA is evaluated from Eq. \ref{PhotonCW_TPA_CW} as

\begin{equation}
	Rate_{TPA,cw}=\frac{N_{in}-N_{out}}{N_{in} \cdot L}=\frac{1-10^{-Loss_{TPA} \cdot L /10}}{L}
	\label{Rate_TPA_CW}
\end{equation}

Assuming that the probability of a TPA transition involving two pulse photons is equal to that of a transition involving one pulse photon and one CW photon, the rate of pulse-photon loss is twice the rate of CW-photon loss. This is because, in the mixed (CW + pulse) interaction, only one CW photon is absorbed per transition, whereas in the pulse–pulse interaction, two pulse photons are absorbed in each TPA event. Therefore,  the rate of photon loss in the pulse due to TPA is calculated as

 \begin{equation}
 \begin{array}{l}
 		Rate_{TPA}=2 \cdot Rate_{TPA,cw}=
 	2 \cdot  \frac{1-10^{-Loss_{TPA} \cdot L /10}}{L}
 \end{array}
 	\label{Rate_TPA}
 \end{equation}

The rate of absorbed pulse photons $Rate_{FCA}$ required to generate the free-carriers corresponding to the measured free-carrier-induced loss $Loss_{FCA}$ is calculated as follows. The number of free carriers was evaluated using well-calibrated relationships between free-carrier absorption and carrier density in silicon nanowire waveguides, as obtained from studies of silicon modulators  (Ref. \cite{SiModulatorIntel2005}).

\begin{equation}
	Loss_{FCA}=8.5\cdot 10^{-18} N_e +6.0\cdot 10^{-18} N_h
	\label{freeCarrierHE}
\end{equation}

where $N_e$ and $N_h$ is the numbers of free electrons and holes in $cm^{-3}$ unit, $Loss_{FCA}$ is the FCA loss measured as shown in Fig. \ref{fig:FigLossTPApFCA} in $cm^{-1}$ unit

Since each TPA event generates one electron–hole pair, the densities of free electrons and free holes are equal. Because two photons are absorbed in each TPA transition, the number of absorbed photons   required to generate the observed free-carrier density is estimated from Eq. \ref{freeCarrierHE} as
 
 \begin{equation}
 	Photon_{FCA}=2 \cdot \dfrac{Loss_{FCA}}{(8.5+6.0)10^{-18}}
 	\label{freeCarrierPhoton}
 \end{equation}
 
 The corresponded rate of photon loss due to FCA is calculated as 
 
    \begin{equation}
   	Rate_{FCA}=\dfrac{Photon_{FCA}}{Photon_{pulse}}
   	\label{RateFCA}
   \end{equation}
   where is $Photon_{pulse}$ is the number of photons in the pulse


\begin{figure}[tb]
	\begin{center}
		\includegraphics[width=7.5 cm]{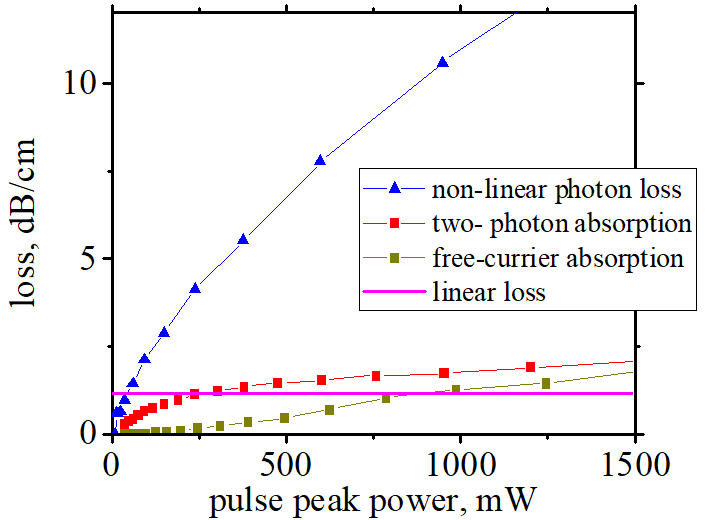}
	\end{center}
	\caption{
		{Three types of nonlinear optical loss in a Si nanowire waveguide: (1) total nonlinear loss induced by a single pulse (blue line), evaluated using Eq. \ref{loss_photon}; (2) loss due to two-photon absorption (red line) evaluated as shown in Fig. \ref{fig:FigLossTPApFCA} by green line; and (3) loss due to free-carrier absorption (yellow line) evaluated as shown in Fig. \ref{fig:FigLossTPApFCA} by red line in comparison with linear loss (pink line).
		} 
	}
	\label{fig:LossAll} 
\end{figure}

\section{Results and Discussion}

\subsection{Non-linear loss}
Figures \ref{fig:LossAll}  compare the linear loss with the three types of nonlinear losses in the Si waveguide. It is important to note that the nonlinear losses are significantly larger than the linear loss, even at relatively small pulse peak powers ($\sim$100 mW). Given their magnitude, nonlinear losses must be a critical consideration in the design of high-speed photonic circuits.

The three types of nonlinear loss impose different limitations on data transmission. Although their magnitudes differ, each affects signal propagation in a distinct way.

The nonlinear loss associated with a single pulse is the largest. However, it behaves similarly to linear loss: it does not accumulate over time and does not introduce cross-talk between channels. It becomes problematic mainly in very long Si waveguides or when pulses with different amplitudes are transmitted.

The loss due to two-photon absorption is smaller, but it is particularly important in dense wavelength-division multiplexing (DWDM) systems. Because two-photon absorption occurs only during the pulse, pulses from different DWDM channels that are temporally aligned can interact through TPA and experience significant cross-talk.

The loss due to free-carrier absorption is the smallest of the three losses, but it accumulates over relatively long timescales ($\sim$100 ns). The loss shown in Fig. \ref{fig:LossAll} corresponds only to free carriers generated by a single 13 ps pulse. Due to carrier accumulation from successive pulses, this type of loss can become substantially larger and may pose a serious limitation for dense and high-speed data processing.

Reducing any of these three types of loss would immediately extend the achievable speed, density, and functional complexity of silicon photonic circuits.

\subsection{Dynamics of photon- electron interaction}

To trace the dynamics of the photon- electron interaction during two-photon absorption, we evaluate a common parameter- the photon loss rate- at each stage of the process. This parameter is expected to be conserved throughout the entire interaction pathway. In an ideal TPA process, two photons are absorbed for each electron transition from the valence band to the conduction band, resulting in the generation of one electron- hole pair. Specifically, the number of nonlinearly absorbed photons should be twice the number of electron transitions from the valence to the conduction band and should correspond directly to the number of generated free carriers. Consequently,   the photon loss rate derived from total nonlinear absorption  should equal the rate inferred from the measured TPA transitions, which in turn should match the rate required to generate the measured free-carrier density.

\begin{figure}[tb]
	\begin{center}
		\includegraphics[width=7 cm]{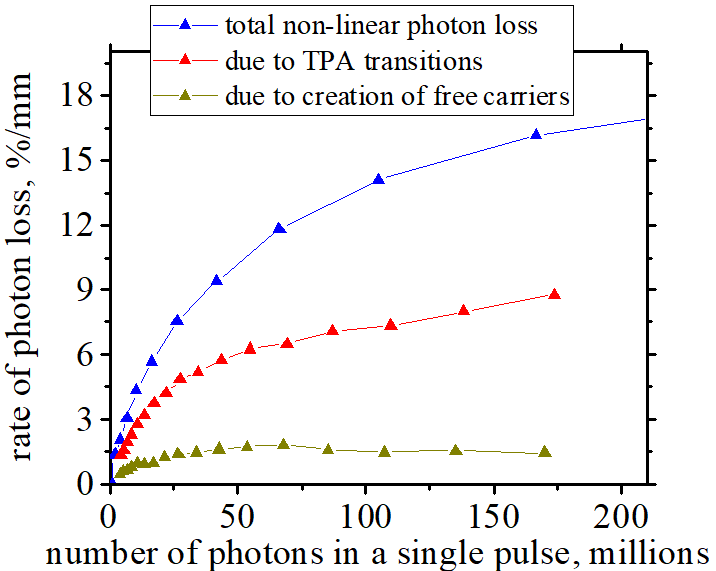}
	\end{center}
	\caption{
		{Rate of photon loss independently measured at three stages of the two-photon absorption process vs. number of photons in a single pulse inside Si: (blue line) nonlinear photon loss of a single pulse; (red line) photon loss due to TPA transitions; (yellow line) photon loss required to generate the measured free carriers.
		} 
	}
	\label{fig:RateAll} 
\end{figure}

Figure \ref{fig:RateAll} compares the photon loss rates as a function of the number of photons in the pump pulse for the three stages of the TPA mechanism. Surprisingly, the three rates differ substantially, and their saturation behaviors are also different.

The largest rate corresponds to the nonlinear absorption of the pulse itself (blue line), which only begins to saturate at a photon number of 200 millions photons per pulse. The rate associated with the TPA transitions (red line) is roughly two times smaller and becomes nearly fully saturated at about 50 millions photons per pulse. The rate required to produce the measured free-carrier density is the smallest and reaches saturation at 20 millions photons per pulse.

Since TPA is a quantum process that generates a discrete number of electron–hole pairs through the absorption of a discrete number of photons, a direct comparison between the number of generated carriers and the number of absorbed photons provides valuable insight into the underlying quantum dynamics of TPA. Figure \ref{figSuppl:FreeElectrons} in Appendix \ref{AppenFreeEl} shows the evaluated density of photoexcited free electrons. The measured free-carrier density is relatively low, despite the substantial nonlinear optical loss observed in the waveguide.

\subsection{Indication of Additional Mechanisms for Non-linear Absorption}

Such large discrepancies among the three measured rates, which should be identical according to the conventional TPA model, indicate that the nonlinear processes in silicon waveguides are more complex than described by the established picture of TPA (Fig. \ref{fig:Dynamics}) and strongly suggest the involvement of additional nonlinear mechanisms or alternative transition pathways.

We do not find a satisfactory alternative explanation. Except, the difference between the red and blue curves in Fig. \ref{fig:RateAll} may be partly attributed to the different probabilities of nonlinear interactions involving two pulse photons compared with those involving one pulse photon and one CW photon. This difference may arise from the significant disparity in coherence properties between the pulse and CW photons.

In principle, measurement 1 and measurement 2 (see Fig. 1) probe the same nonlinear loss mechanism. However, in measurement 1 the loss results from the interaction of two photons within the pulse, whereas in measurement 2 it originates from the interaction between a CW photon and a pulse photon. We interpret these two measurements as corresponding to different steps of the TPA process because the second interaction involves two non-coherent photons, independent of the small difference in their optical frequencies (see Fig. \ref{figSuppl:TPAStatistics}). Such an interaction necessarily proceeds as a two-step process, requiring an intermediate state within the bandgap. In contrast, the interaction between two coherent photons from the same pulse can occur via a simultaneous, one-step transition (e.g. via interband tunneling) and does not require a real intermediate state.

Additionally, some of the observed differences in photon absorption rates (Fig. \ref{fig:RateAll}) may arise from the fact that the measurement of nonlinear loss for a single pulse is influenced by free carriers generated by preceding pump pulses (see Appendix \ref{proceedingPulses}), whereas the measurements of TPA and FCA are not affected by this contribution. However, given that the FCA loss is substantially smaller than the total nonlinear loss of a single pulse  (see Fig. \ref{fig:LossAll}), this particularity of measurements cannot account for the large discrepancy between the TPA loss and the total nonlinear loss (red and blue curves in Fig.~\ref{fig:RateAll}).

\begin{figure}[tb]
	\begin{center}
		\includegraphics[width=7 cm]{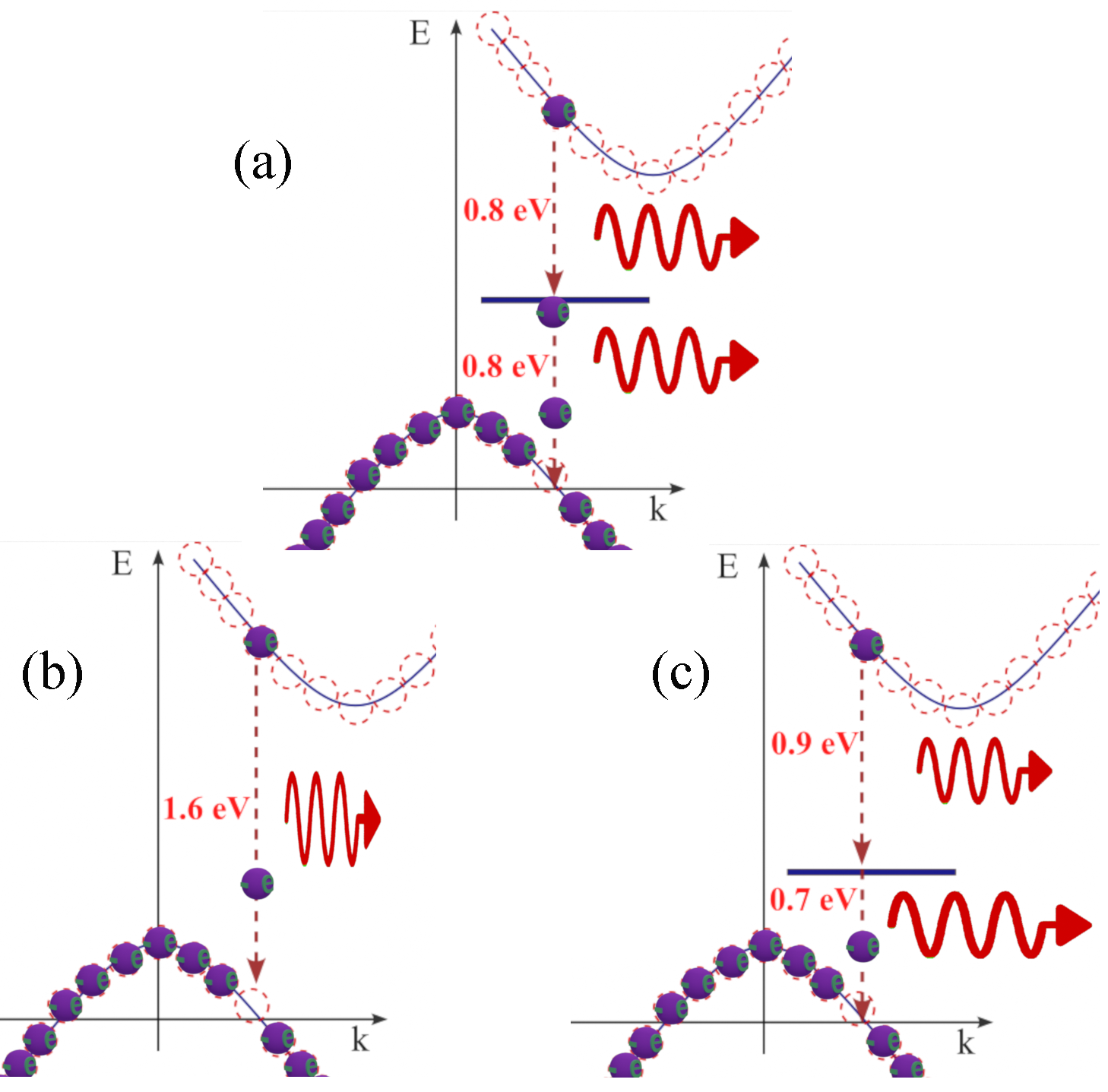}
	\end{center}
	\caption{
		{Possible high-speed relaxation mechanisms of free carriers: (a) reverse TPA, where an electron transitions through the same mid-gap virtual level as in the direct TPA process, emitting two identical photons; (b) direct recombination accompanied by emission of a photon with twice the frequency (second-harmonic emission); and (c) recombination via a different mid-gap virtual level, resulting in the emission of two photons with different energies.
		} 
	}
	\label{fig:FigReverseTPA} 
\end{figure}

Relaxation for free carries is 70 fs Fig.6 of Ref \cite{HotSi2019}

According to theoretical calculations, hot electrons in silicon interact with both acoustic and optical phonons on a characteristic timescale of approximately 10 fs \cite{HotPhotonElectron2022}.

\subsection{Indication of Additional Mechanisms for Electron-Hole Relaxation}

The photon loss rate required to generate the measured free carriers (yellow line in Fig. \ref{fig:RateAll}) is substantially smaller than both the TPA rate (red line) and the total nonlinear photon loss rate (blue line). This discrepancy directly indicates the presence of additional relaxation mechanisms for free carriers beyond the indirect recombination process illustrated in  Fig. \ref{fig:Dynamics}.

Because the number of free carriers was measured immediately after the passage of a 13 ps pulse, any additional relaxation mechanisms must be very fast, with characteristic relaxation times shorter than 13 ps. This suggests that these processes are likely based on direct electronic transitions.

The relaxation time of hot electrons excited approximately 1.6 eV above the valence-band maximum is about 70 fs, as measured by two-photon photoemission \cite{HotSi2019}. This relaxation process involves both energy and momentum relaxation \cite{HotSi2019}. Therefore, direct back-transitions from the conduction band to the valence band may occur within this timescale.

According to theoretical calculations, hot electrons in silicon interact with both acoustic and optical phonons on a characteristic timescale of approximately 10 fs (Ref. \cite{HotPhotonElectron2022}). Following the first electron–phonon scattering event, the excited electron loses its phase coherence to the photons,  which induce its transition, and any subsequent recombination process is not longer coherent.

Therefore, coherent direct back-transitions from the conduction band to the valence band may occur during the first $\sim$10 fs after excitation, while incoherent direct transitions may take place from $\sim$ 10 fs to $\sim$ 70 fs after the electron is excited. As discussed in Section \ref{SectionLossLinear}, in the case of a coherent back-transition, the electron excitation and relaxation cycle does not contribute to optical loss, , because the entire excitation–relaxation cycle is reversible and the absorbed optical energy is returned to the electromagnetic field.

Figure \ref{fig:FigReverseTPA} illustrates three possible fast relaxation mechanisms. The most probable mechanism is the reverse of the TPA process (Fig. \ref{fig:FigReverseTPA}(a)), in which an electron relaxes through the same mid-gap virtual level as in the direct TPA transition, emitting two identical photons.

The existence of this reverse transition is highly plausible because, irrespective of the physical origin of the virtual level, the TPA process does not involve any apparent symmetry breaking that would forbid the corresponding reverse transition. Consequently, the reverse process should be allowed by the same fundamental principles that permit the forward TPA transition.

Another possible mechanism is direct recombination accompanied by the emission of a photon with twice the double frequency (Fig. \ref{fig:FigReverseTPA}(b)). Such a transition is not fundamentally forbidden and therefore may also contribute to the relaxation process.

It should be noted that second-harmonic generation can also occur as a result of interband tunneling \cite{HighGarmonic2018}. However, the underlying physical mechanisms are fundamentally different. In the present case, the emitted second-harmonic photon originates from the direct radiative recombination of an electron–hole pair, whereas in the tunneling-induced process the second-harmonic signal arises from the nonlinear motion of carriers driven by a strong optical field (See details in the next section).

 A third possibility is recombination through a different mid-gap virtual level, leading to the emission of two photons with different energies (Fig. \ref{fig:FigReverseTPA}(c)).

All three relaxation mechanisms compete with the rapid scattering of electrons (and holes) toward the bottom of the conduction band (and the top of the valence band). Once an electron scatters out of the state of the energy or momentum that allows direct transitions, the relaxation mechanisms shown in Fig. \ref{fig:FigReverseTPA} can no longer occur. This time threshold for the direct transitions is expected to be above 10 fs and below 70 fs \cite{HotSi2019,HotPhotonElectron2022}.

Since TPA followed by any of the three  mechanisms shown in Fig. \ref{fig:FigReverseTPA} results in nonlinear optical loss but does not produce long-lived free carriers, these pathways provide a plausible explanation for the large discrepancy between the smaller measured number of  free carriers  (yellow curves in Figs. \ref{fig:LossAll} and \ref{fig:RateAll}) and the much larger number of carriers excited by TPA (blue and red curves). 

\subsection{Two-Photon Absorption vs. Electron Interband Tunneling}
\label{sectionTPAvsKFR}

In two-photon absorption (TPA), an electron undergoes an optical quantum transition from the valence band to the conduction band through a virtual intermediate level as a result of the interaction with and absorption of two photons. However, there exists another mechanism by which an electron can be transferred from the valence band to the conduction band: interband tunneling. Unlike TPA, interband tunneling does not involve an optical quantum transition between discrete energy states.

Interband tunneling can occur in the presence of a sufficiently strong electric field. The electric field tilts the energy bands, causing the electron energy to vary with electron spatial  position. At a sufficiently large field, the energy of the valence-band maximum at one spatial location can become equal to the energy of the conduction-band minimum at another location. If the separation between these locations becomes sufficiently small, an electron can tunnel directly from the valence band to the conduction band conserving its own energy. This phenomenon is known as the Zener effect and forms the operating principle of the Zener diode, a device widely used in electronics from DC to microwave frequencies.

An analogous effect exists in the optical regime, where the strong oscillating electric field of light can induce interband tunneling. This process is commonly described by the Keldysh-Faisal-Reiss (KFR) theory \cite{KFR1964keldysh,KFR1973Faisal,KFR1980Reise}. Because the optical field oscillates on an extremely short timescale- for example, the optical period at a wavelength of 1.55 $\mu m$ is only 5.17 fs- the tunneling process must occur within a half of this period. Consequently, the tunneling distance must be extremely short, on the order of an atomic spacing ($\sim $ 1 Å). For silicon, with a bandgap of 1.12 eV, the electric field required to achieve such tunneling\cite{KFR2016,multi2013}  should be approximately 1 V/Å.

A key distinction between TPA and interband tunneling is that tunneling is driven by the total optical electric field rather than by the absorption of individual photons. In this picture, the transition results from the interaction of an electron with the combined electric field of the optical wave, rather than from two successive photon–electron interactions. Therefore, the process is not constrained by the photon energy. Interband tunneling can occur even when the combined energy of two photons is smaller than the bandgap. Certainly, interband tunneling does not require the existence of a virtual mid-gap level.

Moreover, in contrast to TPA, the probability of interband tunneling generally increases with increasing optical wavelength. At longer wavelengths, the optical field oscillates more slowly, providing a longer time window for tunneling to occur during each optical cycle and thereby increasing the tunneling probability. For this reason, long-wavelength radiation is often preferred for observing strong-field tunneling phenomena. An additional advantage of longer wavelengths is a lower risk of thermal damage, as the photon energy is smaller.  To achieve the high optical intensities required for tunneling while minimizing thermal damage to the material, ultrashort optical pulses, typically shorter than 100 fs, are commonly employed.

In our experiments, the optical pulse duration is relatively long (12.5 ps), and the damage threshold of the silicon waveguides corresponds to a pulse peak power of approximately 2–4 W. Under these conditions, the peak optical electric field does not exceed approximately $6 \cdot 10^{-5}$ V/Å, which is several orders of magnitude smaller than the field strength typically required to induce significant interband tunneling. Therefore, the contribution of interband tunneling is expected to be negligible.

Nevertheless, the possible influence of the KFR mechanism cannot be completely excluded. It is therefore important to distinguish between interband tunneling and two-photon absorption. A key difference is that TPA is governed primarily by the photon number in a pulse, whereas the KFR effect depends directly on the peak electric-field strength (or peak pulse  intensity). Consequently, studying the nonlinear absorption as a function of pulse duration \cite{PulseDuration} could help distinguish between the contributions of these two mechanisms. Unfortunately, the pulse duration of our optical source was fixed, and such measurements could not be performed in the present study.

Another possible way to distinguish between the contributions of TPA and interband tunneling is through the functional dependence of the nonlinear loss on the optical pulse intensity. For two-photon absorption, the nonlinear loss is proportional to the pulse intensity $I$. More generally, for an (n)-photon absorption process, the loss scales as $I^{n-1}$. 

In contrast, the dependence associated with interband tunneling is fundamentally different. The tunneling probability depends on both the thickness of the tunneling barrier and the applied electric field. For a sufficiently thick barrier, the barrier width decreases linearly with increasing electric field. Since the tunneling probability depends exponentially on the barrier width, the transition probability is expected to increase exponentially with the optical electric field. This behavior is also predicted by the KFR model and would result in a sharp increase in nonlinear loss above a certain threshold intensity. Such threshold-like behavior was not observed in our measurements.

However, the barrier width cannot decrease indefinitely in a solid-state material. At a minimum, it should remain comparable to the size of an atom. The periodic atomic lattice is responsible for the formation of the electronic band structure, and at least several atomic layers are required to preserve the separation between the valence and conduction bands. In the limit of a thin barrier with approximately constant width, the tunneling current is expected to be proportional to the applied voltage. Consequently, the rate of electron transfer from the valence band to the conduction band would be proportional to the optical electric field, corresponding to a square-root dependence on the pulse intensity.

The blue curves in Figs. \ref{fig:LossAll} and \ref{fig:RateAll} exhibit a dependence that resembles a square-root function, which could, at first glance, be interpreted as evidence for a KFR contribution. We believe this resemblance is coincidental, because the optical electric field in our experiments is far below the level required to induce significant interband tunneling. A more likely explanation is the saturation of two-photon absorption. At pulse peak powers below approximately 300 mW, the measured dependence is nearly linear, whereas at higher powers the nonlinear loss gradually saturates toward a constant value.

It should also be noted that interband tunneling cannot contribute to Measurement 2 (red curves in Figs. \ref{fig:LossAll} and \ref{fig:RateAll}), which originates from the interaction between the pump pulse and the CW probe photons. The reason is that the pump pulse and CW probe are mutually incoherent. Consequently, the weak electric field of the CW probe does not, on average, modify the much stronger electric field of the pump pulse. Therefore, the KFR mechanism cannot explain the observed signal in Measurement 2.

\subsection{Saturation of Non-Linear Effects}

An interesting feature of the nonlinear behavior of the two-photon absorption process is its saturation at higher optical powers. Surprisingly, all three stages of the TPA mechanism saturate at different photon densities. This indicates that the dynamics of each stage are influenced by distinct mechanisms with different power-dependent behaviors.  Understanding the origin of each saturation effect may help clarify the physical processes governing the different stages of TPA.

One possible explanation for the saturation is the filling of available states in the conduction band. As the number of electronic transitions increases, more states in the conduction band become occupied, reducing the probability of additional TPA transitions. Figure \ref{figSuppl:FreeElectrons} shows the measured density of photoexcited free electrons, as evaluated from Eq.\ref{freeCarrierHE}.  Even at the highest photon densities used in our experiments, the electron density remains relatively low and does not exceed the electron density typically found in moderately or lightly doped silicon, such as that used in a p–n junction. Such a relatively low carrier density is unlikely to cause significant filling of the conduction-band states and therefore cannot explain the strong saturation observed in all stages of the TPA process.

Another possible origin of the saturation is the finite density of states associated with the mid-bandgap virtual level involved in the TPA transition. Depending on the physical origin of this virtual level, the saturation mechanism may differ. If the virtual level is induced by the combined oscillating electric field of the incident photons, then saturation may result from the limited ability of the optical field to generate additional field-induced states. At low and moderate photon numbers, the density of states associated with such field-induced mid-bandgap states may increase approximately in proportion to the optical field strength (or photon number). However, at higher photon densities, further modification of the band structure ceases to increase the state density, leading to saturation of the nonlinear absorption.

Alternatively, if the virtual level corresponds to a conventional mid-gap electronic state with a very short lifetime, saturation could result from a reduction in available unfilled virtual states. As the rate of excitation increases, a larger fraction of the virtual states may remain occupied before the electron either transitions to the conduction band (via TPA) or relaxes back to the valence band. Even if the lifetime of the electron in the virtual state is short, it is not infinitesimally small. Therefore, at sufficiently high excitation rates, the number of available unoccupied virtual states may decrease, leading to saturation of the TPA process.

The faster saturation of the free-carrier generation rate compared with the TPA rate (see Fig. \ref{fig:RateAll}) indicates that a larger population of TPA-excited electrons increases the probability of the relaxation processes illustrated in Fig. \ref{fig:FigReverseTPA}. The detailed physical mechanisms responsible for this behavior remain to be clarified.

Saturation is a well-known characteristic of TPA and has been experimentally observed in the nonlinear changes of both optical loss \cite{PETRA_2021,pulse_NEC_2005} and refractive index  \cite{multiNSatur2001,ZnSe2021}.

\section{Conclusion}

For the first time, we have experimentally traced the complete dynamical pathway of the photon–electron interaction involved in two-photon absorption in a silicon nanowire waveguide. Using three independent high-speed measurements, we evaluated the rates of nonlinear photon absorption, TPA transitions, and free-carrier generation. According to the conventional model of TPA, these three rates should be identical. However, our measurements reveal a significant discrepancy between them. In particular, half of the total nonlinear photon absorption substantially exceeds the number of measured TPA transitions. This observation suggests either the presence of an additional source of nonlinear photon absorption or an intrinsic peculiarity of the TPA process itself. Our results therefore indicate that the TPA mechanism in silicon waveguides is more complex than described by the currently accepted model.

The conventional TPA model assumes that all electrons photoexcited to the conduction band scatter down to the bottom of the conduction band. In contrast, our results show that the number of photoexcited free electrons accumulated at the conduction-band minimum represents only a small fraction of the electrons initially excited by the TPA process. This indicates that a substantial portion of TPA-excited electrons rapidly recombine to the valence band through direct transitions on a timescale shorter than 13 ps.

Furthermore, the three stages of the process exhibit different saturation behaviors, providing additional evidence that the TPA dynamics cannot be fully explained by the classical model.

A major obstacle to understanding the TPA mechanism is the unclear physical origin of the virtual midgap level. We identify three physically plausible origins for this level and discuss how our measurements are consistent with each scenario. However, the present data are not sufficient to determine its true nature, and further investigations are required to clarify the origin of the virtual level underlying the physics of TPA in silicon nanowire waveguides.

Finally, we demonstrate that the nonlinear optical loss associated with TPA is not negligible and can significantly impact the performance of silicon photonic circuits, particularly at high processing speeds and high data densities. We identified three types of nonlinear losses associated with TPA, each affecting signal transmission in a different way. Even at pulse peak powers as low as ~50 mW, the nonlinear loss exceeds the linear propagation loss in silicon waveguides. Therefore, nonlinear effects must be carefully considered in the design of silicon photonic circuits, especially when long waveguides are used. A deeper understanding of the physical mechanisms governing TPA may be essential for mitigating these limitations and improving the performance of future silicon photonic systems.

\section{Acknowledgment}

The authors gratefully acknowledge fruitful discussions with Dr. M. Pelusi.

\bibliography{TwoPhotonAbsorption}

\begin{figure}[tb]
	\begin{center}
		\includegraphics[width=6 cm]{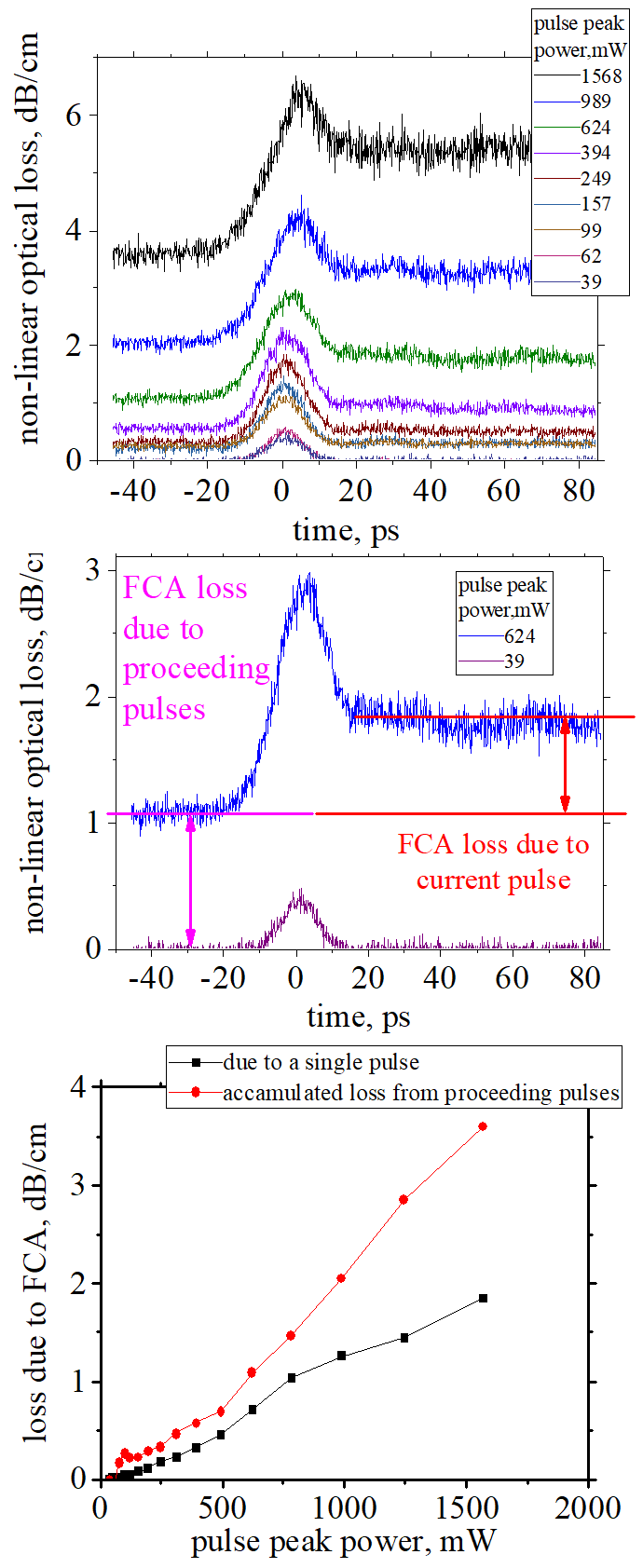}
	\end{center}
	\caption{
		{(a) Temporal dynamics of the nonlinear loss evaluated from the data in Fig. \ref{fig:Exp2}(b) using Eq. \ref{loss_dynamics} without subtraction of $Loss_{before}$ (b) Evaluation of FCA losses due to proceeding pulses and due to current pulse from non-linear loss of CW light. (c) Comparison of FCA losses due to  proceeding  and due to current pulses
		} 
	}
	\label{figSuppl:AccamulatedFCA} 
\end{figure}

\appendix

\section{Loss due to  free-carries created by the proceeding pump }

\label{proceedingPulses}

In our experimental setup, the repetition period of the pulse laser is 52 ns, which is not sufficiently long to allow complete relaxation of the free carriers. As a result, some free carriers excited by preceding pulses remain in the waveguide before the arrival of the next pump pulse. This leads to an initial optical loss that is present even before the pump pulse arrives.

 Figure \ref{figSuppl:AccamulatedFCA}(a) shows the temporal dynamics of the nonlinear loss, similar to those shown in Fig. \ref{fig:Exp2}(c), but without subtraction of the initial loss $Loss_{before}$. As can be seen, the initial loss caused by carriers generated by preceding pulses increases with increasing pulse power.
 
 Figure  \ref{figSuppl:AccamulatedFCA}(b) illustrates how the free-carrier absorption (FCA) losses due to the preceding pulses and due to a single pulse are evaluated independently. Figure  \ref{figSuppl:AccamulatedFCA}(c) compares these two FCA contributions as a function of the pulse peak power. The FCA loss caused by preceding pulses is larger than that produced by the current pulse, and the difference increases at higher pulse powers.

Although the loss induced by the preceding pulse is undesirable, its contribution can be eliminated, as it is done in Eq. (\ref{loss_dynamics}). Consequently, its effect on the present measurements is minimal.

\begin{figure}[tb]
	\begin{center}
		\includegraphics[width=8.5 cm]{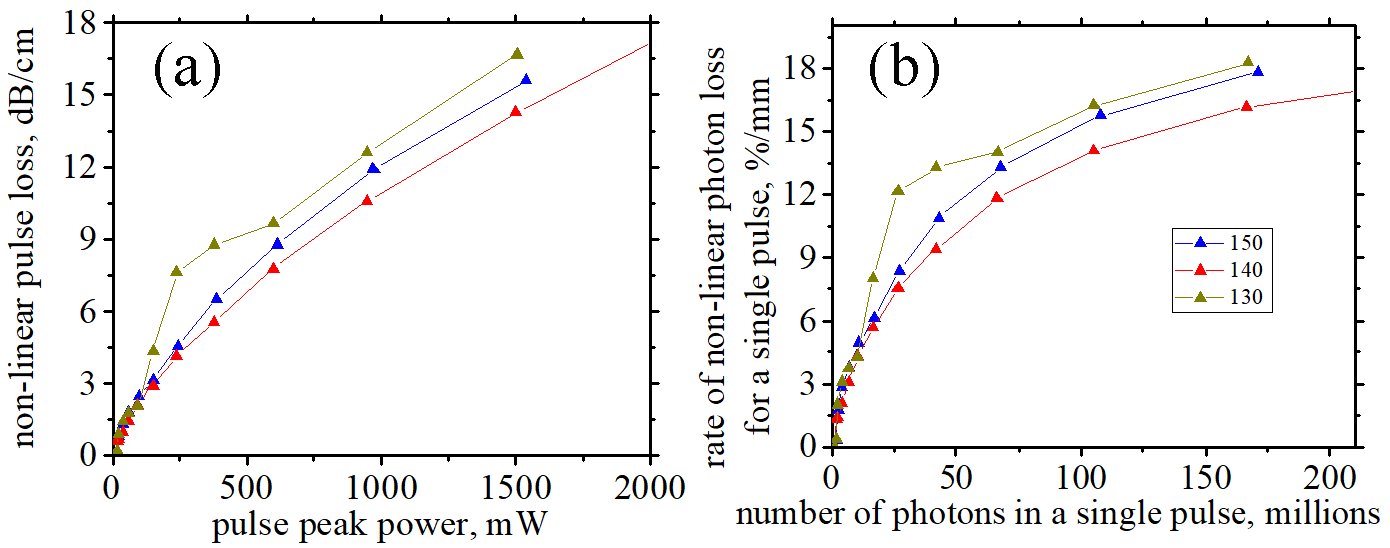}
	\end{center}
	\caption{
		{(a) Nonlinear optical loss and (b) photon loss rate for a single pulse measured at $\lambda_{pulse}$ =1553 nm for waveguides with spot-size converter widths of 130, 140, and 150 nm.
		} 
	}
	\label{figSuppl:PulseStatistics} 
\end{figure}

\section{Reliability and Reproducibility}

The main challenge in reliably measuring nonlinear effects in Si nanowire waveguides is ensuring high stability and repeatability of all insertion losses in the measurement setup. Without such control, the pulse intensity inside the Si waveguide cannot be evaluated accurately, which may introduce substantial errors.

To address this, our setup employs a fully computer-controlled automatic fiber-to-waveguide alignment system. This ensures minimal coupling loss and reproduces the coupling efficiency with a precision of 0.1 dB.

\begin{figure}[tb]
	\begin{center}
		\includegraphics[width=8.5 cm]{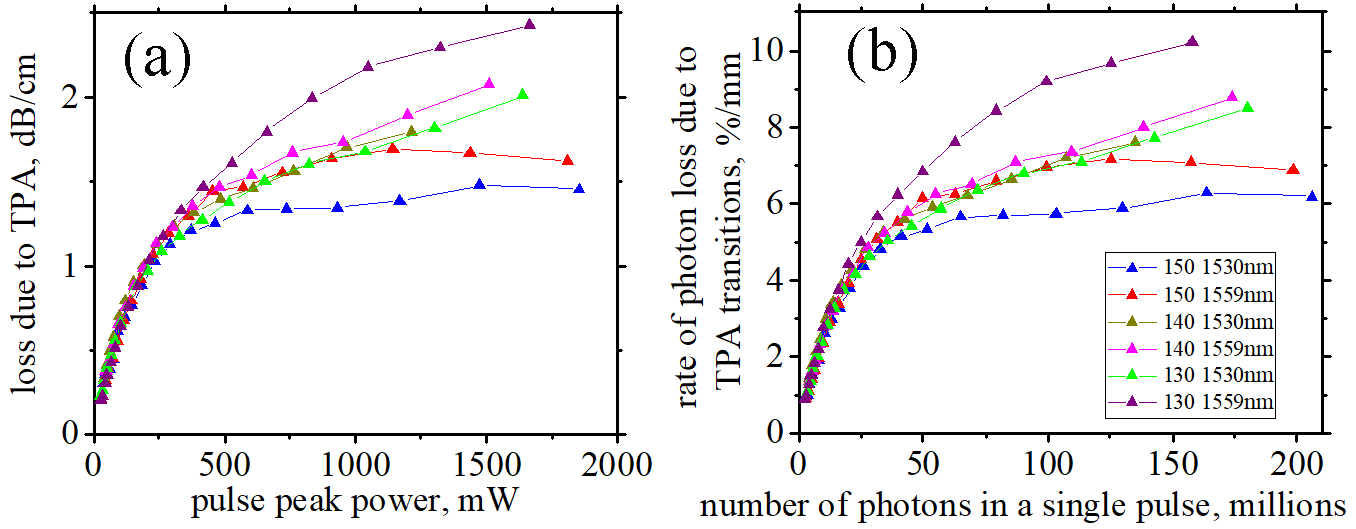}
	\end{center}
	\caption{
		{(a) Nonlinear optical loss and (b) photon loss rate due to TPA  measured at $\lambda_{CW}$ =1530 nm and 1559 nm for waveguides with spot-size converter widths of 130, 140, and 150 nm.
		} 
	}
	\label{figSuppl:TPAStatistics} 
\end{figure}

The robustness of our approach is supported by the consistency of the results: nearly identical measurements were obtained across different waveguides, at various wavelengths, and for samples fabricated at different times.

\begin{figure}[tb]
	\begin{center}
		\includegraphics[width=8.5 cm]{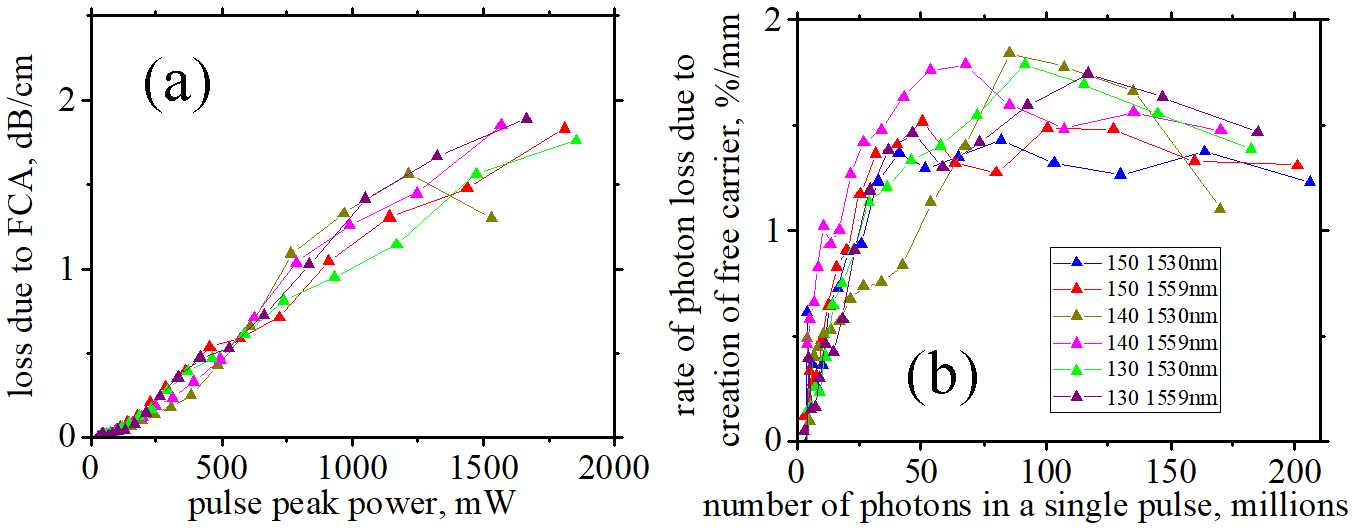}
	\end{center}
	\caption{
		{(a) Nonlinear optical loss and (b) photon loss rate due to FCA  measured at $\lambda_{CW}$ =1530 nm and 1559 nm for waveguides with spot-size converter widths of 130, 140, and 150 nm.
		} 
	}
	\label{figSuppl:FCAStatistics} 
\end{figure}

Figures  \ref{figSuppl:PulseStatistics}, \ref{figSuppl:TPAStatistics} and  \ref{figSuppl:FCAStatistics} compare the nonlinear measurements for waveguides with different spot-size converter widths measured at slightly different wavelengths. In all cases, no dependence of the observed nonlinear effects on either the spot-size converter width or the probe wavelength was observed.

\section{Density of Excited Free Electrons}
\label{AppenFreeEl}

\begin{figure}[tb]
	\begin{center}
		\includegraphics[width=7 cm]{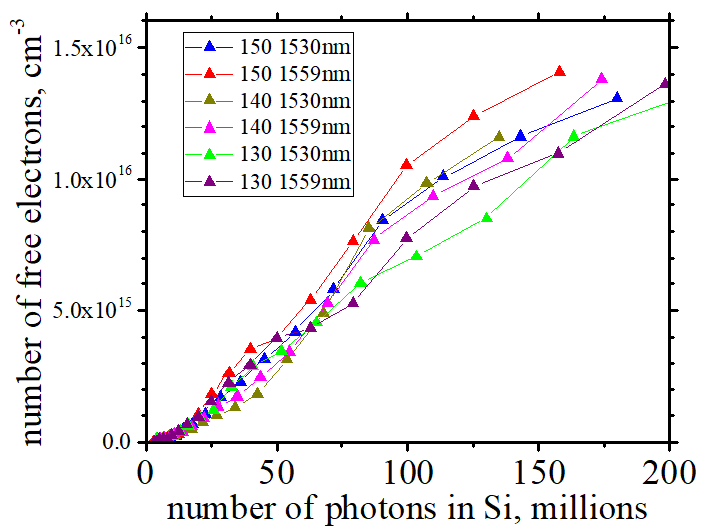}
	\end{center}
	\caption{
		{Density of excited free electrons as a function of the number of photons in the pump pulse inside the Si waveguide.
		} 
	}
	\label{figSuppl:FreeElectrons} 
\end{figure}

Figure \ref{figSuppl:FreeElectrons} shows the measured density of excited free electrons in the Si waveguide. It is important to note that this density remains moderate, even at the highest pulse intensities used, and is comparable to or lower than the electron density in the n-region of a standard Si p–n junction. This suggests that the observed saturation of the nonlinear effects is unlikely to be caused by the filling of available empty states in the conduction band.

\end{document}